\documentclass[aps,a4paper,superscriptaddress,nofootinbib]{revtex4-1}
\usepackage{acronym, amsmath, graphicx, xcolor, ulem, float}
\usepackage[colorlinks,linkcolor=blue,citecolor=blue,urlcolor=blue]{hyperref} 
\floatstyle{plaintop}
\restylefloat{table}
\def\apj{Astrophysical Journal}                
\def\jcap{Journal of Cosmology and Astroparticle Physics}

\def\apjs{Astrophysical Journal, Supplement}              
\def\prd{Phys.~Rev.~D}       
\def\prl{Phys.~Rev.~Lett}    
\def\aap{A\&A}                

\newcommand{\Msun}{\,{\rm M}_\odot}

\begin{document}
\title{Constrains on the electric charges of the binary black holes with GWTC-1 events}

\author{Hai-Tang Wang}
\affiliation{Key Laboratory of Dark Matter and Space Astronomy, Purple Mountain Observatory, Chinese Academy of Sciences, Nanjing 210023, P. R. China.}
\affiliation{School of Astronomy and Space Science, University of Science and Technology of China, Hefei, Anhui 230026, P. R. China.}
\email{wanght@pmo.ac.cn}

\author{Peng-Cheng Li}
\affiliation{Center for High Energy Physics, Peking University, No.5 Yiheyuan Rd, Beijing 100871, P. R. China}
\affiliation{Department of Physics and State Key Laboratory of Nuclear Physics and Technology, Peking University, No.5 Yiheyuan Rd, Beijing 100871, P. R. China}

\author{Jin-Liang Jiang}
\affiliation{Key Laboratory of Dark Matter and Space Astronomy, Purple Mountain Observatory, Chinese Academy of Sciences, Nanjing 210023, P. R. China.}
\affiliation{School of Astronomy and Space Science, University of Science and Technology of China, Hefei, Anhui 230026, P. R. China.}

\author{Guan-Wen Yuan}
\affiliation{Key Laboratory of Dark Matter and Space Astronomy, Purple Mountain Observatory, Chinese Academy of Sciences, Nanjing 210023, P. R. China.}
\affiliation{School of Astronomy and Space Science, University of Science and Technology of China, Hefei, Anhui 230026, P. R. China.}

\author{Yi-Ming Hu}
\affiliation{TianQin Research Center for Gravitational Physics and School of Physics and Astronomy, Sun Yat-sen University (Zhuhai Campus), Zhuhai 519082, P. R. China}

\author{Yi-Zhong Fan}
\affiliation{Key Laboratory of Dark Matter and Space Astronomy, Purple Mountain Observatory, Chinese Academy of Sciences, Nanjing 210023, P. R. China.}
\affiliation{School of Astronomy and Space Science, University of Science and Technology of China, Hefei, Anhui 230026, P. R. China.}

\begin{abstract}

Testing black hole's charged property is a fascinating topic in modified gravity and black hole astrophysics. 
In the first Gravitational-Wave Transient Catalog (GWTC-1), ten binary black hole merger events have been formally reported, and these gravitational wave signals have significantly enhanced our understanding of the black hole. 
In this paper, we try to constrain the amount of electric charge with the parameterized post-Einsteinian framework by treating the electric charge as a small perturbation in a Bayesian way. 
We find that the current limits in our work are consistent with the result of Fisher information matrix method in previous works.
We also develop a waveform model considering a leading order charge effect for binary black hole inspiral. 

\end{abstract}

\maketitle

\section{Introduction}\label{sec:intro}

The prominent black hole (BH) no-hair theorems \citep{Hawking:1971vc, PhysRevLett.34.905} imply vacuum astrophysical black hole can be described by a special case of Kerr-Newman metric, which could be characterized by its mass, spin and charge only \citep{Israel_CMP1967, Carter_PRL1971, Cardoso_IOP2016}. 
Although astrophysical BHs are usually considered as electrically neutral due to charge neutralization by astrophysical plasma, quantum discharge effects, and electron-positron pair production \citep{1969ApJ_Goldreich, Gibbons1975, Pulsars_APJ_Ruderman1975}, an accurate upper limit on the amount of the charges of BHs is still absent. 
Therefore, it is important and meaningful to give a quantitative measurement on the amount of charges of BHs. 
Besides, many novel charging mechanisms has been discussed theoretically, such as primordial black hole \citep{Liu_arxiv2020}, and BHs formed by millicharge dark matter \citep{Cardoso_JCAP2016}. 
In the later case, the discharge process is much slower than for ordinary plasma so the charged BHs are viable. 
So it is meaningful to study the effect on gravitational wave if electric charges are presented in these special theories. 

At present, using the shadow of supermassive BH to estimate its charge has been developed and widely practiced \citep{Johannsen_PRL2016}, for example Sgr A* and M87* measured by VLBI \citep{SgrA_VLBI_Nature2008, SgrA_VLBI_APJ2018, Zajacek_JPCS2019, M87_EHT_APJL2019}. 
These proposals are either far from being accurate or model-dependent. 
Therefore, developing new technique to model-dependently test the charged property is needed, and gravitational wave has always been expected, which encodes the BH's information \citep{Yunes_PRD2009}. 
Fortunately, after decades of hard work, LIGO-Virgo announced the first detection of the gravitational wave (GW) signal GW150914 in 2015, generated by the merger of a binary BH \citep{gw150914_PRL2016}. 
To characterize GW, we often use inspiral, merger and ringdown to describe the whole coalescence of a binary BH, where the inspiral stage is generally described by post-Newtonian theory \citep{Blanchet_LRR2013}, merger stage is generally approximated by the numerical simulation \citep{Lehner_RAA2014}, and ringdown stage is described by the BH perturbation theory \citep{Teukolsky_APJ1973,Teukolsky_APJ1974}. 
Some waveform models, such as IMRPhenomPv2 (IMR) waveform model, have incorporated these three stages \citep{Hannam_PRL2015}.

The GW waveform will be affected if the two BHs are electrically charged. 
For simplicity, consider the electric dipole radiation during their orbital motions, which will be reflected in the phase of inspiral stage, and thus provides us a new trick to detect charged binary BH. 
Depending on the amount of the charge, there are two different treatments. 
When the charge of the black hole is small, the dipole radiation can be taken as a perturbation or correction to the phase of the inspiral stage of the GW waveform, which can be incorporated into the parameterized post-Einsteinian (ppE) framework \citep{Yunes_PRD2009}. 
In this case, the correction due to the electric dipole radiation is completely described by the coefficient of the $-1$ post-Newtonian (PN) order in the waveform, the rest part of the waveform is the same as that of the accurate waveform model describing the coalescence of two neutral black holes, such as the well-known phenomenological waveform model \citep{Hannam_PRL2015}. 
This is similar to the dipole radiation caused by the scalar charges carried by black holes in the scalar-tensor theory \citep{Arun_PRD2012}. 
However, if the charge of the black hole is large, the ppE framework may not be applicable anymore, as the latter requires the expansion in PN is always linear, which may not be assured as a prior. 

We also consider only the leading order gravitational quadrupole radiation and the electric dipole radiation, based on which we call such waveform the leading order charged (LOC) one. 
It is apparent that the LOC waveform has a bad accuracy when applied to the analysis of GW data unless the inspiral duration is long enough and the orbit of the two black holes decays very slow. 
Despite this shortcoming, the LOC waveform provides us a toy model such that the effect of the charge can be  demonstrated explicitly without  the assumption that the charge of the black hole is small. 
The detailed calculation of the LOC waveform is shown in Sec.~\ref{sec:LOC}. 

In this work, based on IMR waveform model in LIGO-Virgo Algorithm Library \citep{lalsuite}, we use the Bayesian method to test the dipole radiation of GW signals \citep{Yunes_PRD2016, Test_GR_PRD2019} with the ppE framework. 
As the main conclusion, we find no visible charge taking by astrophysical BH.

The rest of this paper is organized as follows. 
In Sec.~\ref{sec:effects}, we introduce the ppE framework. 
In Sec.~\ref{sec:method}, we briefly introduce the Bayesian method for the GW data processing and the results are presented in Sec.\ref{sec:result}. 
In order to verify the consistency  of the results obtained from ppE method, in Sec.~\ref{sec:LOC}, we introduce the toy model, the LOC waveform, to study the effect of the amount of the charge on the waveform. 
In Sec.~\ref{sec:summary}, we summarize the calculation results and present the conclusion, and discuss the deficiencies of this work as well as what can be done further.
We assume $c = 1$ throughout the paper unless otherwise specified.

\section{Effects on gravitational wave signals}\label{sec:effects}

In this section, we introduce the waveform models adopted in this work. 
We treat the charge effect as a small perturbation, which can be well described by the parameterized post-Einsteinian framework. 
In this case, the phase of the inspiral part of the charged binary BHs can be incorporated into the ppE formalism, which describes the gravitational waveforms of theories alternative to general relativity (GR) in a model-independent way \citep{Yunes_PRD2009, Yunes_PRD2016}. 
Formally, the effect of the electric dipole radiation can be captured by the ppE parameter entering at $-1$ post-Newtonian order, similar to the dipole radiation from the scalar-tensor theory \citep{Arun_PRD2012}. 

Assuming the early-inspiral stage of the IMR waveform model is reduced to the form of $h_{\mathrm{e-ins}}(f)=A_{\mathrm{e-ins}}(f)e^{i{\Psi}_{\mathrm{e-ins}}}$, then the waveform model due to modifications from different modified gravity effects can be written as 
\begin{equation}\label{eq:modify_insp}
h_{\mathrm{e-ins}}(f)=A_{\mathrm{e-ins}}(f)e^{i{\Psi}_{\mathrm{e-ins}}+i\Delta\Psi}
\end{equation}
In ppE framework \citep{Yunes_PRD2016, Test_GR_PRD2019}, 
\begin{equation}\label{eq:ppE}
\Delta\Psi_{\rm ppE}=\beta(\pi G\mathcal{M}f)^{b/3}, 
\end{equation}
where $\beta$ is the amplitude coefficient and $b$ is the exponent coefficient. 
Note that the modification enters at $(b+5)/2$ PN order for a related modified gravity. 
$\mathcal{M}=(m_1m_2)^{3/5}/(m_1+m_2)^{1/5}$ is the chirp mass with component masses $m_1$ and $m_2$, and $f$ is the frequency of the GW. 

The study in \citet{Tahura_PRD2019} shows that at least in theories where the leading corrections enter at negative PN orders, the phase-only analyzes can produce sufficiently accurate constraints. 
In this analysis, we only consider the correction on the phase and neglect the correction to the amplitude. 
The frequency at the end of this stage is given by $f_c=0.018/[G(m_1+m_2)]$ \citep{Test_GR_PRD2019}, above which this model is calibrated with numerical-relativity data and can not be applied to ppE formalism. 

The effect of the charge on the waveform can be taken as a perturbation term in the phase \citep{Cardoso_JCAP2016, Christiansen_arxiv2020}, 
\begin{equation}\label{phase}
\Delta\Psi_q=-\frac{5}{3584}\eta^{2/5}\zeta^2\kappa^2\left(\pi\kappa G\mathcal{M}f\right)^{-7/3}
\end{equation}
where $\eta=m_1m_2/(m_1+m_2)^2$ is symmetric mass ratio, $\zeta$ represents the difference between the charges of the binary BH and is defined by $\zeta=|\lambda_1-\lambda_2|/\sqrt{1-\lambda_1\lambda_2}$, where $\lambda_i=q_i/(\sqrt{G}m_i)$ is the charge-to-mass ratio and $q_i$ is the electric charge of a BH. 
Here $\kappa=1-\lambda_1\lambda_2\approx 1$ since we consider the effects of $\lambda_1$ and $\lambda_2$ as small perturbations. 
In this case, $\beta=-\frac{5}{3584}\eta^{2/5}\zeta^2$ and $b=-7$, corresponding to Eq.~(\ref{eq:ppE}). 

In TABLE I. of \citet{Yunes_PRD2016}, Yunes et al. obtained constraint $|\beta|\lesssim 1.6\times 10^{-4}$ at $-1$ PN order for GW150914. 
The relationship between $\beta$ and $\zeta$ is: $\beta=-5/3584\eta^{2/5}\zeta^2$. 
Thus, one can derive the corresponding constraint $\zeta\lesssim 0.45$. 
There are two ways to enhance the constraint on the charge. 
Firstly, one may expect the increase of the sensitivities of GW detectors, and a more stringent constraint on the charge should appear. 
Secondly, a full gravitational waveform model of charged binary BH is excepted to give a more stringent constraint on the charge. 
This full GW waveform should apply to the coalescence of two black holes carried with arbitrary amount of charge. 
It is a challenging job and in this work we move forward by developing the leading order charged (LOC) waveform model, which is introduced in Sec.~\ref{sec:LOC}. 

\section{Bayesian inference methods}\label{sec:method}

To infer the uncertainty of the source parameters $\vec{\mathcal{\theta}}$, which are quantified by the posterior probability distribution $p(\vec{\mathcal{\theta}}|d, M)$, we perform Bayesian analysis with the prior $p(\vec{\mathcal{\theta}}|M)$ and the likelihood with Gaussian noise assumption for the GW data, 
\begin{equation}
\label{eq:likelihood}
\begin{aligned}
p(d|\vec{\mathcal{\theta}},M)\propto \mathrm{exp}\left[-\frac{1}{2}\sum^{N}_{i=1}\langle d_i-h_i|d_i-h_i\rangle\right],
\end{aligned}
\end{equation}
where $d_i$ is the data of the $i$-th instrument, $M$ is the model assumption, $N$ is the number of detectors in the network of Advanced LIGO and Advanced Virgo, and $h_i$ is the waveform model calculated with $\vec{\mathcal{\theta}}$ for the $i$-th detector. 
The noise weighted inner product $\langle a(f)|b(f)\rangle$ is defined by 
\begin{equation}
\label{eq:inner_product}
\langle a(f)|b(f)\rangle = 4\mathcal{R}\int^{f_{high}}_{f_{low}}\frac{a(f)b(f)}{S_n(f)} \mathrm{d}f, 
\end{equation}
where $f_{low}$ and $f_{figh}$ are the high and low pass cut-off frequencies respectively, $S_n(f)$ is the power spectral density of the detector noise. 
The Bayesian theorem is described by 
\begin{equation}
\label{eq:bayes}
p(\vec{\mathcal{\theta}}|d,M) = \frac{p(d|\vec{\mathcal{\theta}},M)p(\vec{\mathcal{\theta}}|M)}{p(d|M)},
\end{equation}
where $p(d|M)=\int p(d|\vec{\mathcal{\theta}},M)p(\vec{\mathcal{\theta}}|M)\mathrm{d}\vec{\mathcal{\theta}}$ is the evidence of a specific model assumption. 

There are fifteen parameters in the IMR waveform model, including the redshifted chirp mass ($\mathcal{M}_z$), mass ratio ($q$), luminosity distance ($d_L$), inclination angle ($\theta_{jn}$), the reference orbital phase ($\phi_c$), the geocentric time ($t_c$), the polarization angle ($\psi$), the two dimensional sky location, and six spin parameters. 
And the early-inspiral stage of the IMR waveform model for ppE framework (hereafter e-insp-ppE waveform model) has an additional ppE parameter $\zeta$. 
We marginalize over the reference phase $\phi_c$ and the geocentric time $t_c$, thus we have $14$ free parameters for e-insp-ppE waveform model. 
The GW data and power spectral density for each event are downloaded from LIGO-Virgo GW Open Science Center \citep{LIGO_O1O2data_arxiv2019}. 
To estimate parameters with data from the first two observation runs (O1 and O2), we carry out Bayesian inference with {\sc Bilby} \citep{Ashton_APJ2019}, using {\sc Pymultinest} \citep{Buchner_AAP2014} as our sampler. 

The prior on the redshifted chirp mass is chosen to be uniform in the range of $5\Msun \leq\mathcal{M}_z\leq 20\Msun$ for GW151226 and GW170608 while $5\Msun \leq\mathcal{M}_z\leq 50\Msun$ for the other events, the prior on the mass ratio is chosen to be uniform in the range of $0.25\leq\mathrm{q}\leq 1$. 
We apply comoving uniform prior on the luminosity distance $50\mathrm{Mpc}\leq d_L\leq 4000\mathrm{Mpc}$, while the prior on the inclination angle is chosen to be uniform in the range of $-1\leq\cos\iota\leq 1$. 
The prior on the polarization angle is chosen to be uniform in the range of $0\leq\psi\leq \pi$, and the prior of the sky location is chosen to be uniform in spherical coordinates. 
The prior on $|\zeta|$ is chosen to be uniform in the range of $0\leq|\zeta|<\sqrt{2}$. 
For other parameters in the e-insp-ppE waveform model, we use the same priors presented in \citet{LIGO_PRX2019}. 

\section{Results of Bayesian inferences}\label{sec:result}

In this section, on the promise that the BH has a small amount of charge, we set up constraints on the dipole radiation with the ppE framework. 
We analyze with e-insp-ppE waveform model to constrain the dipole radiation, with a high frequency cut-off at $f_c=0.018/[G(m_1+m_2)]$. 
Specifically, this analysis is applied to GW150914, GW151226, GW170104, GW170608, and GW170814. 
We neglect other events since the early-inspiral signal-to-noise ratios (SNRs) of them are all less than $6$. 

\begin{table}[htb!]
        \begin{tabular}{@{\extracolsep{4pt}}c|c|c|c|c|c@{}}
            \hline
            \hline
            Event name&GW150914&GW151226&GW170104&GW170608&GW170814\\
            \hline
            $f_c[\rm Hz]$& $50$& $153$& $60$& $179$& $58$ \\
            \hline
            $|\zeta|$ & $0.20_{-0.17}^{+0.22}$ & $0.16_{-0.10}^{+0.10}$ & $0.38_{-0.29}^{+0.32}$ & $0.14_{-0.07}^{+0.07}$ & $0.12_{-0.10}^{+0.15}$ \\
            \hline
            \hline
        \end{tabular}
    \caption{The constraints on dipole radiation of five selected binary BH events, $90\%$ confidence intervals are shown. 
    }\label{table:dipole}
\end{table}

As shown in Table \ref{table:dipole}, for GW170608, the GW data can constrain $|\zeta|$ to be less than about $0.21$ at $90\%$ credible level, while the loosest case is given by GW170104, $|\zeta|\leq 0.70$ (at $90\%$ credible level) due to its lowest SNR of the inspiral signal among all these five events. 
In \citet{Yunes_PRD2016}, the authors got constraints on scalar dipole radiation through a Fisher parameter estimation study, using a fitted spectral noise sensitivity curve. 
As a result, the constraint on GW150914 is $|\zeta|\lesssim 0.45$, and the constraint on GW151226 is $|\zeta|\lesssim 0.24$. 
Their results agree well with ours shown in Table \ref{table:dipole}. 
One can also get constraints on $|\zeta|$ by reweighting the posteriors of parameters $\delta\phi_{-2}$ from results in \citet{Test_GR_PRD2019}, where $\delta\phi_{-2}=-5/84\zeta^2$ is the parameterized violation of GR at $-1$ PN. 
Recently, a similar analysis was performed in \citet{2021arXiv210407590W} with both reweighting method and Bayesian inference method, and they find that the Bayesian analysis is more reliable than the reweighting analysis. 

\begin{figure}[htb]
\centering
\includegraphics[width=0.8\linewidth]{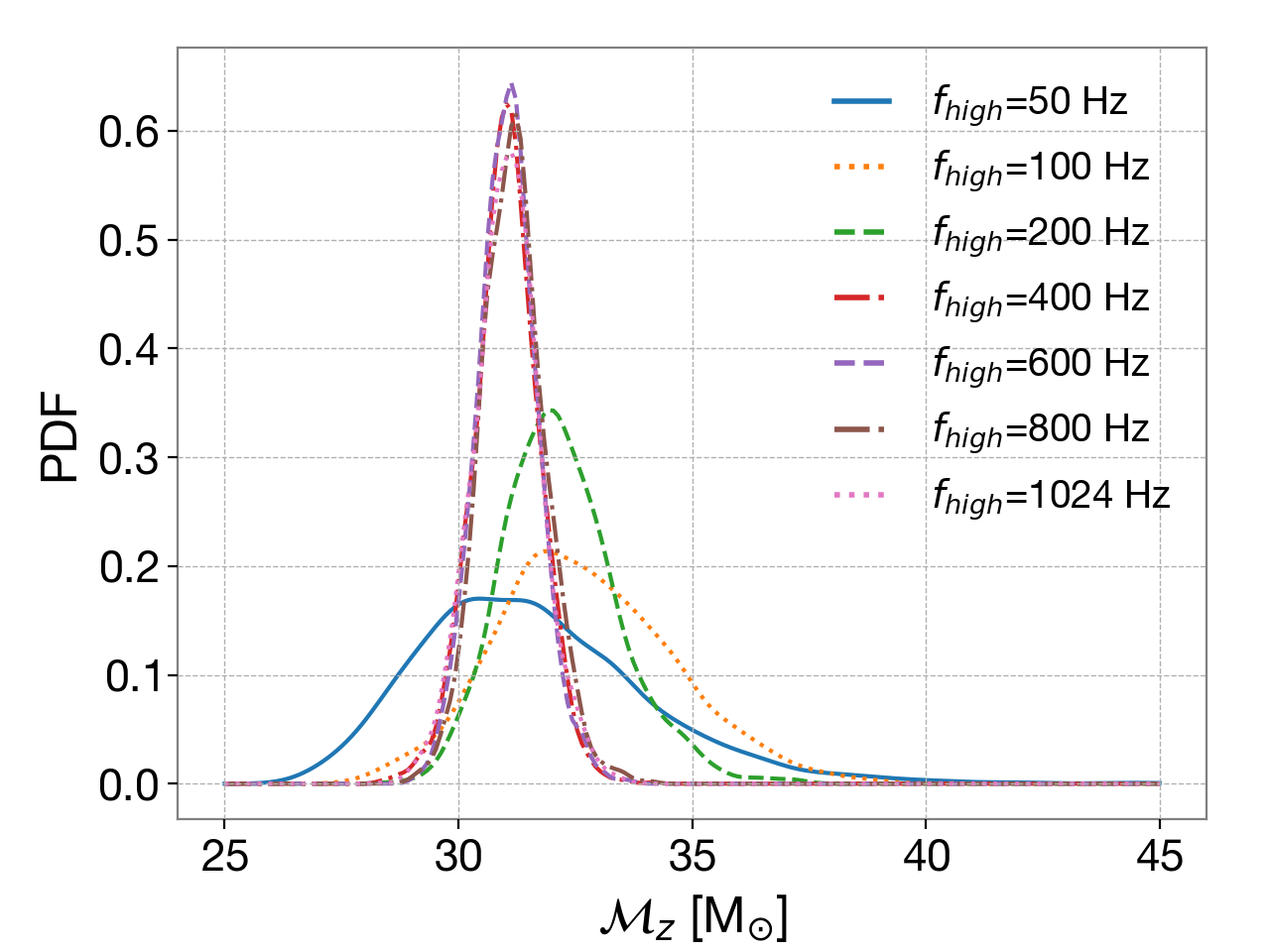}
\caption{The marginalized posterior probability distributions of chirp-mass of GW150914, with different high frequency cut-off $f_{high}=50,100,200,400,600,800$, and $1024$ Hz. 
The uncertainty of the distribution becomes larger as $f_{high}$ is less than $400$ Hz.}
\label{fig:high_cuts}
\end{figure}

To further discuss the reliability of our results, we apply Bayesian inference on GW data of GW150914 with a set of low-pass cut-off frequencies $f_{high}$. 
And we show the chirp mass distribution in Figure \ref{fig:high_cuts}, in which the distribution becomes wider as $f_{high}$ less than $400$ Hz. 
Nevertheless, the result is not biased even the cut-off frequency is as low as $50$ Hz, although the uncertainty is a bit larger. 

Actually, the charge will significantly affect the space-time metric of the binary system if the charge is as large as the maximum allowable value. 
This effect will be fully reflected in waveform model when higher PN order terms are taken into account. 
However, an analytic solution will not be available in the forthcoming future and one can only get it numerically, i.e. \citet{2021PhRvL.126d1103B}. 
In \citet{2021PhRvL.126d1103B}, the binary charged black holes were studied via numerical general relativity method, where the higher PN order effects of the charge are included and the amount of charge is set free. 
One of the main conclusions from this paper is that the greatest difference between charged and uncharged black holes arises in the earlier inspiral. 
This provides a convincing evidence to support the reasonability of our work.

For future space-based GW detectors such as Laser Interferometer Space Antenna (LISA) \citep{LISA_arxiv2017} and TianQin \citep{TQ_2015}, the duration of GW strain from BH binaries are much longer \citep{Liu_PRD2020, WangPRD2019}. 
Since the dipole gravitational radiation dominates at $-1$ PN order, it is more suitable for detecting by LISA and TianQin. 
\citet{Barausse_PRL2016} shows that joint observations of GW150914-like systems by LIGO-Virgo and LISA will improve bounds on dipole emission from BH binaries by several orders of magnitude relative to current constraints. 
We expect that with multi-band GW the constraint on the electric charge of BH can be improved as well. 

\section{The Leading Order Charged Waveform}\label{sec:LOC}

In this section, we try to learn more about the effect of the charge by developing a LOC waveform model. 
For this waveform model, we do not consider the spin of binary black holes since the spin evolution of the binary BH only occurs at high orders of the PN expansion \citep{Kidder:1992fr, Cutler:1994ys}. 
Therefore, we only need to analyze the orbit (circular) evolution in the inspiral phase due to the energy loss. 
For two point particles with mass $m_1$ and $m_2$ and charge $q_1$ and $q_2$ respectively, we define $\lambda_i=q_i/(\sqrt{G}m_i)$. 
They orbit each other with orbital radius $R$, and the orbit decays with time $t$. 
The dissipation of total energy can be divided into two parts, one is the emission of GW, the other is electromagnetic dipole radiation, which can be written as \citep{Cardoso_JCAP2016}

\begin{equation}\label{eq:energy_evolution_CBH}
\begin{aligned} 
\frac{d E}{d t} =-\frac{d E_{\mathrm{GW}}}{d t}-\frac{d E_{\mathrm{dip}}}{d t} =-\frac{32}{5 } \eta^{2} \frac{GG_{\mathrm{eff}}^{3} M^{5}}{R^{5}}-\frac{2}{3} \zeta^{2} \frac{G_{\mathrm{eff}}^{3} m_{1}^{2} m_{2}^{2}}{R^{4}} \ ,
\end{aligned}
\end{equation}
where $M=m_1+m_2$ is total mass of the binary, $G_{\mathrm{eff}}=G(1-\lambda_1\lambda_2)$ is the effective Newton constant, $\zeta=|\lambda_1-\lambda_2|/\sqrt{1-\lambda_1\lambda_2}$ is used to characterize the difference between the charges of two BHs, and $\eta=m_1m_2/(m_1+m_2)^2$ is symmetric mass ratio. 
The evolution equation of the orbital radius arising from Eq.(\ref{eq:energy_evolution_CBH}) is

\begin{equation}
\label{eq:orbit_evolution} 
-\frac{d R(t)}{d t}=\frac{A}{R(t)^{3}}+\frac{B}{R(t)^{2}} \ ,
\end{equation}
where $A=64GG_{\mathrm{eff}}^{2} M m_{1} m_{2}/5$ and $B=4 \zeta^{2} G_{\mathrm{eff}}^{2} m_{1} m_{2}/3$. 
When $B$ is not equal to zero, the relation between $R$ and $t$ can be parameterized as 
\begin{equation}\label{eq:t_r_analytic}
\begin{aligned}
\tau&=\frac{1}{3B}R^3-\frac{A}{2B}R^2+\frac{A^2}{B^3}R-\frac{A^3}{B^4}\ln\left(\frac{B}{A}R+1\right) \\
&\simeq\frac{R^4}{A}\left(\frac{1}{4}-\frac{B}{5A}R+\frac{B^2}{6A^2}R^2-\frac{B^3}{7A^3}R^3+\frac{B^4}{8A^4}R^4\right) +\mathcal{O}\left(\frac{BR}{A}\right)^9 
\end{aligned}
\end{equation}
where $\tau=t_c-t$ and $t_c$ is the coalescence time. 
It is clear that an analytical solution of $R(t)$ is not possible in Eq. (\ref{eq:t_r_analytic}) , so we have to solve it numerically. 
Due to the last part of Eq. (\ref{eq:t_r_analytic}), the numerical solution is not accurate enough when $BR/A\sim 0$, thus we expand Eq. (\ref{eq:t_r_analytic}) as a series of $BR/A$ . 

According to the Kepler's law, the orbital frequency is $\sqrt{G_{\mathrm{eff}}M/R^3}$ and the gravitational frequency is twice as much as the orbital frequency, so we have $\omega_{gw}=2\pi f_{\mathrm{gw}}=2\sqrt{G_{\mathrm{eff}}M/R^3}$. 
The waveform in time domain contains two parts (\citep{Cutler:1994ys,Maggiore_1900}) 
\begin{equation}\label{eq:time_domain_charge}
\begin{aligned} 
h_{+}(t)=&\frac{4}{d_L}\left(\frac{G_{\mathrm{eff}} \mathcal{M}}{c^{2}}\right)^{5 / 3}\left(\frac{\pi f_{\mathrm{gw}}\left(t\right)}{c}\right)^{2 / 3} 
\frac{1+\cos ^{2} \iota}{2} \cos \left(\Phi\left(t\right)\right) \ ,\\
h_{\times}(t)=&\frac{4}{d_L}\left(\frac{G_{\mathrm{eff}} \mathcal{M}}{c^{2}}\right)^{5 / 3}\left(\frac{\pi f_{\mathrm{gw}}\left(t\right)}{c}\right)^{2 / 3} 
\cos \iota \sin \left(\Phi\left(t\right)\right) \ ,
\end{aligned} 
\end{equation}
where $\iota$ is inclination angle, $d_L$ is luminosity distance and the phase of waveform is
\begin{equation}
\Phi(t+t_{\mathrm{ISCO}})=\int_{t_{\mathrm{ISCO}}}^{t} d t^{\prime} \omega_{\mathrm{gw}}\left(t^{\prime}\right) +\phi_c \ ,
\end{equation}
where $t_{\mathrm{ISCO}}$ is the time when the BH reaches the innermost stable circular orbit (ISCO). 
We cut the phase before $t_{\mathrm{ISCO}}$ for three reasons: first, the phase before $t_{\mathrm{ISCO}}$ can be included in $\phi_c$ and the effect is equal to a time shift, which means this does not affect the results of parameter estimation; second, $\omega_{\mathrm{gw}}$ is infinity when $\tau=0$, and this can not be integrated; last but not least, the LOC waveform model cannot truly describe the motion of the binary BH when the orbital distance is too small. 

According to Eqs. (\ref{eq:orbit_evolution}) and (\ref{eq:t_r_analytic}), $t_{\mathrm{ISCO}}$ is fixed for a given $R_{\mathrm{ISCO}}$. 
In reality, it is difficult to know the exact values of $M$ and $\lambda$ for the final BH. 
Because with a given $\lambda_1\lambda_2$ and $|\zeta|$, we still cannot uniquely determine the respective charge of the two BHs. 
For example if $\lambda_1=a$, $\lambda_2=b$ is the solution then  $\lambda_1=b$, $\lambda_2=a$ could still be the solution. 
Approximately we take $M=m_1+m_2, \lambda=\min\{|\frac{m_1\lambda_1+m_2\lambda_2}{m_1+m_2}|,|\frac{m_1\lambda_2+m_2\lambda_1}{m_1+m_2}|\}$. 
then in the similar way to \citet{Pugliese_PRD2011}  we obtain
\begin{equation}\label{eq:isco_charge}
R_{\mathrm{ISCO}}=\frac{4GM\lambda^2}{3+1/C+C} \ , 
\end{equation}
where $C = -(9-8\lambda^2-4\sqrt{4\lambda^4-9\lambda^2+5})^{1/3}$. 
The ISCO of a charged BH decreases with the charge, and we have $R_{\mathrm{ISCO}}=4GM$ for $|\lambda|=1$ and $R_{\mathrm{ISCO}}=6GM$ for the non-rotating uncharged BH, respectively. 

\begin{figure}[!ht]
\centering
\includegraphics[width=0.45\columnwidth,height=7cm]{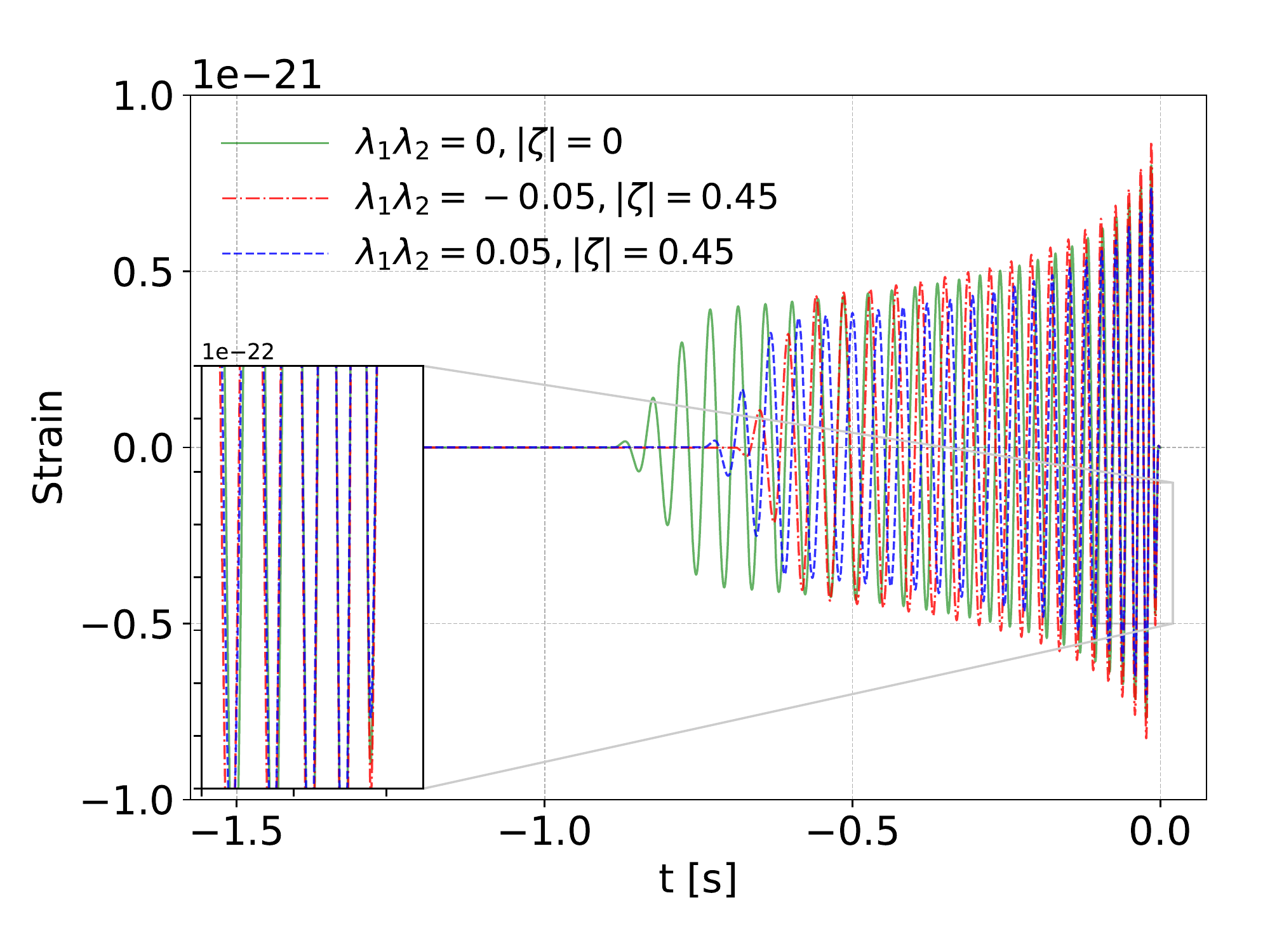}
\includegraphics[width=0.45\columnwidth,height=7cm]{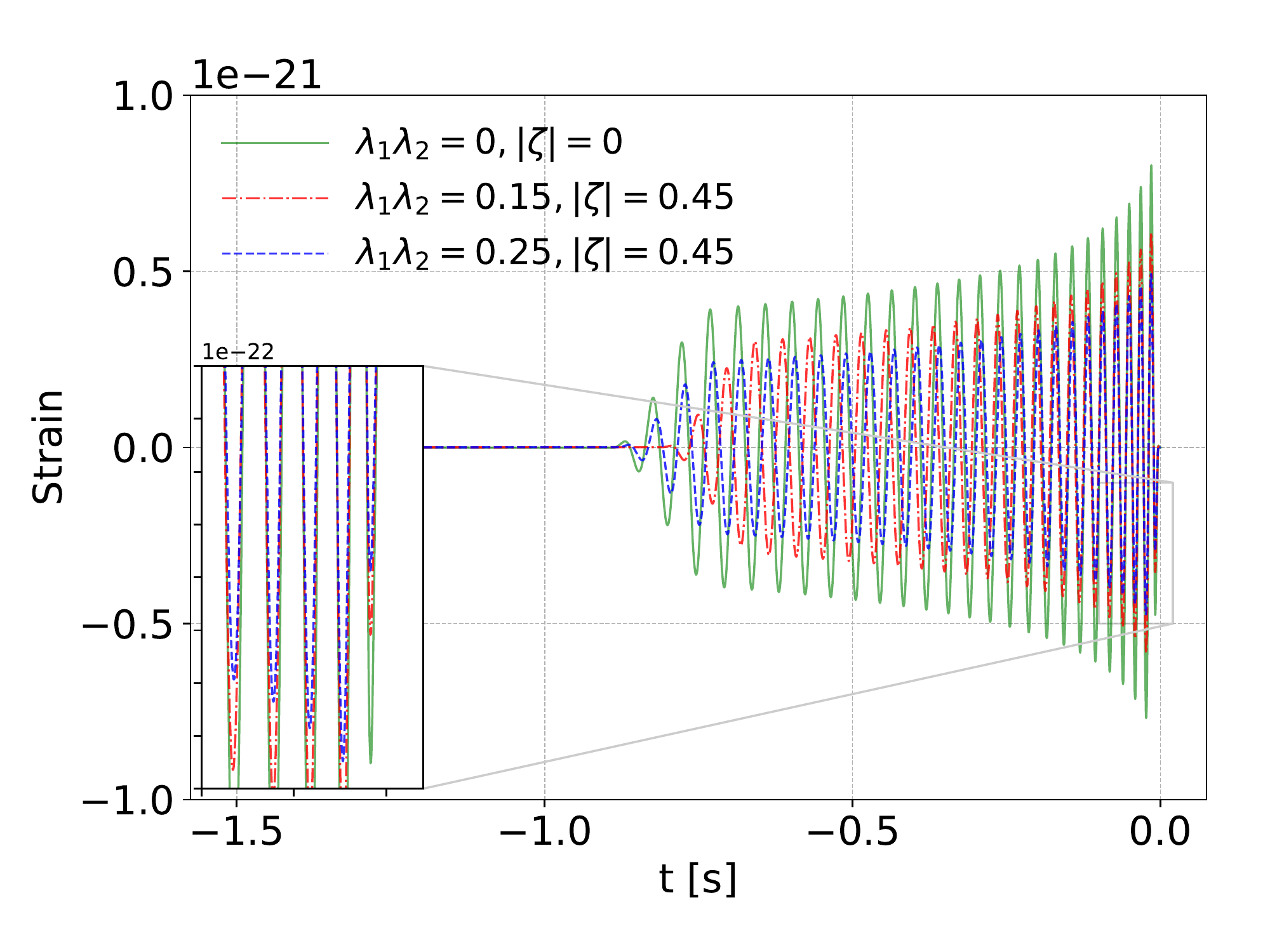}
\caption{
The tapered LOC gravitational waveforms for BH binaries with different charges. 
The rest parameters of these waveforms are the same, i.e., the detector-frame masses are $m_1=m_2=30\Msun$, the luminosity distance is $d_L=540$ Mpc, the inclination angle is $\theta_{jn}=0$. 
The low cut-off frequency is $f_{low}=20$ Hz. 
The green solid line represents the case of zero charge, and the red (blue) dash dotted (dashed) line in the left panel represents the waveform with parameter $\lambda_1\lambda_2=-0.05\ (0.05)$, the red (blue) dash dotted (dashed) line in the right panel represents the waveform with parameter $\lambda_1\lambda_2=0.15\ (0.25)$.
}
\label{fig:0PN_0PNC}
\end{figure}

Based on the discussion above, we obtain the LOC waveform model. 
Obviously, it reduces to the 0PN when both BHs are uncharged. 
As we have already pointed out, under the ppE framework, LIGO-Virgo has very weak restrictions on the charges of GW150914, with $|\zeta|\lesssim 0.45$. 
Here we show that such large amount of charge could have non-negligible effect on the waveform. 
We choose parameters similar to GW150914, where $m_1=m_2=30\Msun,d_L=540\mathrm{Mpc},\iota=0,\phi_c=0$, and we set low cut-off frequency to be $f_{low}=20$ Hz. 
For the charge of GW150914, if $|\zeta|=0.45$, then possible allowed range of $\lambda_1\lambda_2$ is $(-0.053, 0.80)$. 
For proper comparison, we have chosen four special values with $\lambda_1\lambda_2=\{-0.05,0.05,0.15,0.25\}$ to study the effect of charge. 
As shown in Figure \ref{fig:0PN_0PNC}, both the amplitude and phase of the GW signal will be affected significantly if GW150914 is endowed with a large amount of charges. 
The effect of charge on the early inspiral of the gravitational waveform is more significant, which is consistent with previous works \citep{Yunes_PRD2009,Yunes_PRD2016,Cardoso_JCAP2016}. 
It is also worth noting that $G_{\mathrm{eff}}$ and $\mathcal{M}$ have similar effects on shaping the GW signals, the smaller $G_{\mathrm{eff}}$, the lower GW amplitude, as described in Eq. (\ref{eq:time_domain_charge}). 
We also find that the larger value of $\lambda_1\lambda_2$, the slower frequency evolution of GW signals. 

\section{Summary}\label{sec:summary}

In this work, we study the charge effect on the inspiral stage of the GW waveform. 
By considering the charge effect as an perturbation in the inspiral stage, we perform Bayesian inference on five events of the first Gravitational-Wave Transient Catalog (GWTC-1), GW150914, GW151226, GW170104, GW170608, and GW170814, whose SNRs are larger than $6$ after applying a low-frequency cut off $f_c$. 
Based on the ppE formalism, the most stringent limit from FIM up to now is $|\zeta|\sim 0.24$ \citep{Yunes_PRD2016, Cardoso_JCAP2016}, while our Bayesian based result is $|\zeta|\sim 0.21$ at the $90\%$ credible level, which is given by GW170608. 
The analysis on GW151226 and GW170814 give similar constraints, i.e., $\zeta<0.26$ and $\zeta<0.27$ at $90\%$ credible level respectively. 
For GW150914, the constraint is $\zeta<0.42$ at $90\%$ credible level, which is similar to the analysis before \citep{Yunes_PRD2016}. 

To explicitly show the effect of the charge on the waveform, we developed the LOC waveform model in Sec.~\ref{sec:LOC}. 
Although due to the lack of accuracy, this waveform cannot be used to analyze the realistic GW data, this toy model has the advantage of being adjustable and intuitive. 
The LOC waveform is obtained by taking into account the dissipation effect caused by the electric dipole radiation and the quadrupole gravitational radiation on the circular orbit of the charged binary BHs. 
Like the 0PN waveform model, the spin of BH is not considered for the LOC waveform model as which emerges in the waveform only at higher orders. 
Besides, the analytical LOC waveform is achievable only if the charge is treated as a correction, so the general charge case is gotten numerically. 
We find that both the amplitude and phase of the GW signal will be strongly affected if the BHs are endowed with a large amount of charge. 

The work in this paper can be improved in several aspects. 
For example, in the employment of the ppE framework, we only consider the electric dipole radiation, so when the two black holes have the same charge-to-mass ratio, the charge effect disappears in the phase of the ppE waveform. 
To overcome this problem, one can consider the electric quadrupole radiation. 
Moreover, as we have shown above, the constraints on the electric charges of the black holes are still not very stringent, one of the possible ways to improve this is to add higher order corrections to the waveform. 
However, this may not be helpful, because the parameterized constraints on the PN coefficients obtained by LIGO-Virgo show that the -1 PN gets the most stringent constraint than other higher PN orders \citep{Test_GR_PRD2019}, which we expect also applies to the study of the charged black holes. 
The contributions from the higher orders corrections of some other specific theories has been studied in \citet{Yunes_PRD2016}.  

Above all, we give the constraints on the dipole radiations of GW events from GWTC-1. 
These constraints can be also converted to the constraints on radiation effects in other forms, such as the radiation caused by the magnetic or other U(1) dark charges carried by black holes. 
In the future, similar analysis can be applied to GW data that published in GWTC-2 \citep{LIGO_O3a_arxiv2020}. 

\section{Acknowledgments}
We sincerely thank Vitor Cardoso and Laura Bernard for their kindhearted help. 
Hai-Tian Wang also appreciates Yi-Fu Cai, Jian-dong Zhang, Bing-bing Zhang, Si-ming Liu, Can-ming Deng, Ming-Zhe Han, and Shao-Peng Tang for insightful comments and discussions. 
This work has been supported by NSFC under grants of No. 11525313 (i.e., Funds for Distinguished Young Scholars), No. 11921003, No. 11335012, No. 11325522, No. 11735001, No. 11703098, No. 11847241 and No. 11947210. 
Peng-Cheng Li is also funded by China Postdoctoral Science Foundation Grant No. 2020M670010. 
This research has made use of data, software and/or web tools obtained from the Gravitational Wave Open Science Center (https://www.gw-openscience.org), a service of LIGO Laboratory, the LIGO Scientific Collaboration and the Virgo Collaboration. 
LIGO is funded by the U.S. National Science Foundation. Virgo is funded by the French Centre National de Recherche Scientifique (CNRS), the Italian Istituto Nazionale della Fisica Nucleare (INFN) and the Dutch Nikhef, with contributions by Polish and Hungarian institutes.

\bibliographystyle{apsrev4-1}
\bibliography{LIGO_BH_charge}

\begin{thebibliography}{44}%
\makeatletter
\providecommand \@ifxundefined [1]{%
 \@ifx{#1\undefined}
}%
\providecommand \@ifnum [1]{%
 \ifnum #1\expandafter \@firstoftwo
 \else \expandafter \@secondoftwo
 \fi
}%
\providecommand \@ifx [1]{%
 \ifx #1\expandafter \@firstoftwo
 \else \expandafter \@secondoftwo
 \fi
}%
\providecommand \natexlab [1]{#1}%
\providecommand \enquote  [1]{``#1''}%
\providecommand \bibnamefont  [1]{#1}%
\providecommand \bibfnamefont [1]{#1}%
\providecommand \citenamefont [1]{#1}%
\providecommand \href@noop [0]{\@secondoftwo}%
\providecommand \href [0]{\begingroup \@sanitize@url \@href}%
\providecommand \@href[1]{\@@startlink{#1}\@@href}%
\providecommand \@@href[1]{\endgroup#1\@@endlink}%
\providecommand \@sanitize@url [0]{\catcode `\\12\catcode `\$12\catcode
  `\&12\catcode `\#12\catcode `\^12\catcode `\_12\catcode `\%12\relax}%
\providecommand \@@startlink[1]{}%
\providecommand \@@endlink[0]{}%
\providecommand \url  [0]{\begingroup\@sanitize@url \@url }%
\providecommand \@url [1]{\endgroup\@href {#1}{\urlprefix }}%
\providecommand \urlprefix  [0]{URL }%
\providecommand \Eprint [0]{\href }%
\providecommand \doibase [0]{http://dx.doi.org/}%
\providecommand \selectlanguage [0]{\@gobble}%
\providecommand \bibinfo  [0]{\@secondoftwo}%
\providecommand \bibfield  [0]{\@secondoftwo}%
\providecommand \translation [1]{[#1]}%
\providecommand \BibitemOpen [0]{}%
\providecommand \bibitemStop [0]{}%
\providecommand \bibitemNoStop [0]{.\EOS\space}%
\providecommand \EOS [0]{\spacefactor3000\relax}%
\providecommand \BibitemShut  [1]{\csname bibitem#1\endcsname}%
\let\auto@bib@innerbib\@empty
\bibitem [{\citenamefont {Hawking}(1972)}]{Hawking:1971vc}%
  \BibitemOpen
  \bibfield  {author} {\bibinfo {author} {\bibfnamefont {S.~W.}\ \bibnamefont
  {Hawking}},\ }\href {\doibase 10.1007/BF01877517} {\bibfield  {journal}
  {\bibinfo  {journal} {Commun. Math. Phys.}\ }\textbf {\bibinfo {volume}
  {25}},\ \bibinfo {pages} {152} (\bibinfo {year} {1972})}\BibitemShut
  {NoStop}%
\bibitem [{\citenamefont {Robinson}(1975)}]{PhysRevLett.34.905}%
  \BibitemOpen
  \bibfield  {author} {\bibinfo {author} {\bibfnamefont {D.~C.}\ \bibnamefont
  {Robinson}},\ }\href {\doibase 10.1103/PhysRevLett.34.905} {\bibfield
  {journal} {\bibinfo  {journal} {Phys. Rev. Lett.}\ }\textbf {\bibinfo
  {volume} {34}},\ \bibinfo {pages} {905} (\bibinfo {year} {1975})}\BibitemShut
  {NoStop}%
\bibitem [{\citenamefont {Israel}(1968)}]{Israel_CMP1967}%
  \BibitemOpen
  \bibfield  {author} {\bibinfo {author} {\bibfnamefont {W.}~\bibnamefont
  {Israel}},\ }\href {\doibase 10.1007/BF01645859} {\bibfield  {journal}
  {\bibinfo  {journal} {Commun. Math. Phys.}\ }\textbf {\bibinfo {volume}
  {8}},\ \bibinfo {pages} {245} (\bibinfo {year} {1968})}\BibitemShut {NoStop}%
\bibitem [{\citenamefont {Carter}(1971)}]{Carter_PRL1971}%
  \BibitemOpen
  \bibfield  {author} {\bibinfo {author} {\bibfnamefont {B.}~\bibnamefont
  {Carter}},\ }\href {\doibase 10.1103/PhysRevLett.26.331} {\bibfield
  {journal} {\bibinfo  {journal} {Phys. Rev. Lett.}\ }\textbf {\bibinfo
  {volume} {26}},\ \bibinfo {pages} {331} (\bibinfo {year} {1971})}\BibitemShut
  {NoStop}%
\bibitem [{\citenamefont {{Cardoso}}\ and\ \citenamefont
  {{Gualtieri}}(2016)}]{Cardoso_IOP2016}%
  \BibitemOpen
  \bibfield  {author} {\bibinfo {author} {\bibfnamefont {V.}~\bibnamefont
  {{Cardoso}}}\ and\ \bibinfo {author} {\bibfnamefont {L.}~\bibnamefont
  {{Gualtieri}}},\ }\href {\doibase 10.1088/0264-9381/33/17/174001} {\bibfield
  {journal} {\bibinfo  {journal} {Classical and Quantum Gravity}\ }\textbf
  {\bibinfo {volume} {33}},\ \bibinfo {eid} {174001} (\bibinfo {year}
  {2016})},\ \Eprint {http://arxiv.org/abs/1607.03133} {arXiv:1607.03133
  [gr-qc]} \BibitemShut {NoStop}%
\bibitem [{\citenamefont {{Goldreich}}\ and\ \citenamefont
  {{Julian}}(1969)}]{1969ApJ_Goldreich}%
  \BibitemOpen
  \bibfield  {author} {\bibinfo {author} {\bibfnamefont {P.}~\bibnamefont
  {{Goldreich}}}\ and\ \bibinfo {author} {\bibfnamefont {W.~H.}\ \bibnamefont
  {{Julian}}},\ }\href {\doibase 10.1086/150119} {\bibfield  {journal}
  {\bibinfo  {journal} {\apj}\ }\textbf {\bibinfo {volume} {157}},\ \bibinfo
  {pages} {869} (\bibinfo {year} {1969})}\BibitemShut {NoStop}%
\bibitem [{\citenamefont {Gibbons}(1975)}]{Gibbons1975}%
  \BibitemOpen
  \bibfield  {author} {\bibinfo {author} {\bibfnamefont {G.~W.}\ \bibnamefont
  {Gibbons}},\ }\href {\doibase 10.1007/BF01609829} {\bibfield  {journal}
  {\bibinfo  {journal} {Communications in Mathematical Physics}\ }\textbf
  {\bibinfo {volume} {44}},\ \bibinfo {pages} {245} (\bibinfo {year}
  {1975})}\BibitemShut {NoStop}%
\bibitem [{\citenamefont {{Ruderman}}\ and\ \citenamefont
  {{Sutherland}}(1975)}]{Pulsars_APJ_Ruderman1975}%
  \BibitemOpen
  \bibfield  {author} {\bibinfo {author} {\bibfnamefont {M.~A.}\ \bibnamefont
  {{Ruderman}}}\ and\ \bibinfo {author} {\bibfnamefont {P.~G.}\ \bibnamefont
  {{Sutherland}}},\ }\href {\doibase 10.1086/153393} {\bibfield  {journal}
  {\bibinfo  {journal} {\apj}\ }\textbf {\bibinfo {volume} {196}},\ \bibinfo
  {pages} {51} (\bibinfo {year} {1975})}\BibitemShut {NoStop}%
\bibitem [{\citenamefont {{Liu}}\ \emph {et~al.}(2020)\citenamefont {{Liu}},
  \citenamefont {{Guo}}, \citenamefont {{Cai}},\ and\ \citenamefont
  {{Kim}}}]{Liu_arxiv2020}%
  \BibitemOpen
  \bibfield  {author} {\bibinfo {author} {\bibfnamefont {L.}~\bibnamefont
  {{Liu}}}, \bibinfo {author} {\bibfnamefont {Z.-K.}\ \bibnamefont {{Guo}}},
  \bibinfo {author} {\bibfnamefont {R.-G.}\ \bibnamefont {{Cai}}}, \ and\
  \bibinfo {author} {\bibfnamefont {S.~P.}\ \bibnamefont {{Kim}}},\ }\href
  {\doibase 10.1103/PhysRevD.102.043508} {\bibfield  {journal} {\bibinfo
  {journal} {\prd}\ }\textbf {\bibinfo {volume} {102}},\ \bibinfo {eid}
  {043508} (\bibinfo {year} {2020})},\ \Eprint
  {http://arxiv.org/abs/2001.02984} {arXiv:2001.02984 [astro-ph.CO]}
  \BibitemShut {NoStop}%
\bibitem [{\citenamefont {{Cardoso}}\ \emph {et~al.}(2016)\citenamefont
  {{Cardoso}}, \citenamefont {{Macedo}}, \citenamefont {{Pani}},\ and\
  \citenamefont {{Ferrari}}}]{Cardoso_JCAP2016}%
  \BibitemOpen
  \bibfield  {author} {\bibinfo {author} {\bibfnamefont {V.}~\bibnamefont
  {{Cardoso}}}, \bibinfo {author} {\bibfnamefont {C.~F.~B.}\ \bibnamefont
  {{Macedo}}}, \bibinfo {author} {\bibfnamefont {P.}~\bibnamefont {{Pani}}}, \
  and\ \bibinfo {author} {\bibfnamefont {V.}~\bibnamefont {{Ferrari}}},\ }\href
  {\doibase 10.1088/1475-7516/2016/05/054} {\bibfield  {journal} {\bibinfo
  {journal} {\jcap}\ }\textbf {\bibinfo {volume} {2016}},\ \bibinfo {eid} {054}
  (\bibinfo {year} {2016})},\ \Eprint {http://arxiv.org/abs/1604.07845}
  {arXiv:1604.07845 [hep-ph]} \BibitemShut {NoStop}%
\bibitem [{\citenamefont {Johannsen}\ \emph {et~al.}(2016)\citenamefont
  {Johannsen}, \citenamefont {Broderick}, \citenamefont {Plewa}, \citenamefont
  {Chatzopoulos}, \citenamefont {Doeleman}, \citenamefont {Eisenhauer},
  \citenamefont {Fish}, \citenamefont {Genzel}, \citenamefont {Gerhard},\ and\
  \citenamefont {Johnson}}]{Johannsen_PRL2016}%
  \BibitemOpen
  \bibfield  {author} {\bibinfo {author} {\bibfnamefont {T.}~\bibnamefont
  {Johannsen}}, \bibinfo {author} {\bibfnamefont {A.~E.}\ \bibnamefont
  {Broderick}}, \bibinfo {author} {\bibfnamefont {P.~M.}\ \bibnamefont
  {Plewa}}, \bibinfo {author} {\bibfnamefont {S.}~\bibnamefont {Chatzopoulos}},
  \bibinfo {author} {\bibfnamefont {S.~S.}\ \bibnamefont {Doeleman}}, \bibinfo
  {author} {\bibfnamefont {F.}~\bibnamefont {Eisenhauer}}, \bibinfo {author}
  {\bibfnamefont {V.~L.}\ \bibnamefont {Fish}}, \bibinfo {author}
  {\bibfnamefont {R.}~\bibnamefont {Genzel}}, \bibinfo {author} {\bibfnamefont
  {O.}~\bibnamefont {Gerhard}}, \ and\ \bibinfo {author} {\bibfnamefont
  {M.~D.}\ \bibnamefont {Johnson}},\ }\href {\doibase
  10.1103/PhysRevLett.116.031101} {\bibfield  {journal} {\bibinfo  {journal}
  {Phys. Rev. Lett.}\ }\textbf {\bibinfo {volume} {116}},\ \bibinfo {pages}
  {031101} (\bibinfo {year} {2016})}\BibitemShut {NoStop}%
\bibitem [{\citenamefont {Doeleman}\ \emph {et~al.}(2008)\citenamefont
  {Doeleman} \emph {et~al.}}]{SgrA_VLBI_Nature2008}%
  \BibitemOpen
  \bibfield  {author} {\bibinfo {author} {\bibfnamefont {S.}~\bibnamefont
  {Doeleman}} \emph {et~al.},\ }\href {\doibase 10.1038/nature07245} {\bibfield
   {journal} {\bibinfo  {journal} {Nature}\ }\textbf {\bibinfo {volume}
  {455}},\ \bibinfo {pages} {78} (\bibinfo {year} {2008})},\ \Eprint
  {http://arxiv.org/abs/0809.2442} {arXiv:0809.2442 [astro-ph]} \BibitemShut
  {NoStop}%
\bibitem [{\citenamefont {Lu}\ \emph {et~al.}(2018)\citenamefont {Lu} \emph
  {et~al.}}]{SgrA_VLBI_APJ2018}%
  \BibitemOpen
  \bibfield  {author} {\bibinfo {author} {\bibfnamefont {R.-S.}\ \bibnamefont
  {Lu}} \emph {et~al.},\ }\href {\doibase 10.3847/1538-4357/aabe2e} {\bibfield
  {journal} {\bibinfo  {journal} {Astrophys. J.}\ }\textbf {\bibinfo {volume}
  {859}},\ \bibinfo {pages} {60} (\bibinfo {year} {2018})},\ \Eprint
  {http://arxiv.org/abs/1805.09223} {arXiv:1805.09223 [astro-ph.GA]}
  \BibitemShut {NoStop}%
\bibitem [{\citenamefont {Zaja{\v{c}}ek}\ \emph {et~al.}(2019)\citenamefont
  {Zaja{\v{c}}ek}, \citenamefont {Tursunov}, \citenamefont {Eckart},
  \citenamefont {Britzen}, \citenamefont {Hackmann}, \citenamefont {Karas},
  \citenamefont {Stuchl{\'{\i}}k}, \citenamefont {Czerny},\ and\ \citenamefont
  {Zensus}}]{Zajacek_JPCS2019}%
  \BibitemOpen
  \bibfield  {author} {\bibinfo {author} {\bibfnamefont {M.}~\bibnamefont
  {Zaja{\v{c}}ek}}, \bibinfo {author} {\bibfnamefont {A.}~\bibnamefont
  {Tursunov}}, \bibinfo {author} {\bibfnamefont {A.}~\bibnamefont {Eckart}},
  \bibinfo {author} {\bibfnamefont {S.}~\bibnamefont {Britzen}}, \bibinfo
  {author} {\bibfnamefont {E.}~\bibnamefont {Hackmann}}, \bibinfo {author}
  {\bibfnamefont {V.}~\bibnamefont {Karas}}, \bibinfo {author} {\bibfnamefont
  {Z.}~\bibnamefont {Stuchl{\'{\i}}k}}, \bibinfo {author} {\bibfnamefont
  {B.}~\bibnamefont {Czerny}}, \ and\ \bibinfo {author} {\bibfnamefont {J.~A.}\
  \bibnamefont {Zensus}},\ }\href {\doibase 10.1088/1742-6596/1258/1/012031}
  {\bibfield  {journal} {\bibinfo  {journal} {Journal of Physics: Conference
  Series}\ }\textbf {\bibinfo {volume} {1258}},\ \bibinfo {pages} {012031}
  (\bibinfo {year} {2019})}\BibitemShut {NoStop}%
\bibitem [{\citenamefont {{Event Horizon Telescope Collaborat}}\ \emph
  {et~al.}(2019)\citenamefont {{Event Horizon Telescope Collaborat}},
  \citenamefont {Akiyama} \emph {et~al.}}]{M87_EHT_APJL2019}%
  \BibitemOpen
  \bibfield  {author} {\bibinfo {author} {\bibnamefont {{Event Horizon
  Telescope Collaborat}}}, \bibinfo {author} {\bibfnamefont {K.}~\bibnamefont
  {Akiyama}},  \emph {et~al.} (\bibinfo {collaboration} {Event Horizon
  Telescope}),\ }\href {\doibase 10.3847/2041-8213/ab0ec7} {\bibfield
  {journal} {\bibinfo  {journal} {Astrophysical Journal Letters}\ }\textbf
  {\bibinfo {volume} {875}} (\bibinfo {year} {2019}),\
  10.3847/2041-8213/ab0ec7}\BibitemShut {NoStop}%
\bibitem [{\citenamefont {Yunes}\ and\ \citenamefont
  {Pretorius}(2009)}]{Yunes_PRD2009}%
  \BibitemOpen
  \bibfield  {author} {\bibinfo {author} {\bibfnamefont {N.}~\bibnamefont
  {Yunes}}\ and\ \bibinfo {author} {\bibfnamefont {F.}~\bibnamefont
  {Pretorius}},\ }\href {\doibase 10.1103/PhysRevD.80.122003} {\bibfield
  {journal} {\bibinfo  {journal} {Phys. Rev. D}\ }\textbf {\bibinfo {volume}
  {80}},\ \bibinfo {pages} {122003} (\bibinfo {year} {2009})}\BibitemShut
  {NoStop}%
\bibitem [{\citenamefont {Abbott}\ \emph {et~al.}(2016)\citenamefont {Abbott},
  \citenamefont {Abbott},\ and\ \citenamefont {Abbott}}]{gw150914_PRL2016}%
  \BibitemOpen
  \bibfield  {author} {\bibinfo {author} {\bibfnamefont {B.~P.}\ \bibnamefont
  {Abbott}}, \bibinfo {author} {\bibfnamefont {R.}~\bibnamefont {Abbott}}, \
  and\ \bibinfo {author} {\bibfnamefont {T.~D. e.~a.}\ \bibnamefont {Abbott}}
  (\bibinfo {collaboration} {LIGO Scientific Collaboration and Virgo
  Collaboration}),\ }\href {\doibase 10.1103/PhysRevLett.116.061102} {\bibfield
   {journal} {\bibinfo  {journal} {Phys. Rev. Lett.}\ }\textbf {\bibinfo
  {volume} {116}},\ \bibinfo {pages} {061102} (\bibinfo {year}
  {2016})}\BibitemShut {NoStop}%
\bibitem [{\citenamefont {Blanchet}(2014)}]{Blanchet_LRR2013}%
  \BibitemOpen
  \bibfield  {author} {\bibinfo {author} {\bibfnamefont {L.}~\bibnamefont
  {Blanchet}},\ }\href {\doibase 10.12942/lrr-2014-2} {\bibfield  {journal}
  {\bibinfo  {journal} {Living Rev. Rel.}\ }\textbf {\bibinfo {volume} {17}},\
  \bibinfo {pages} {2} (\bibinfo {year} {2014})},\ \Eprint
  {http://arxiv.org/abs/1310.1528} {arXiv:1310.1528 [gr-qc]} \BibitemShut
  {NoStop}%
\bibitem [{\citenamefont {Lehner}\ and\ \citenamefont
  {Pretorius}(2014)}]{Lehner_RAA2014}%
  \BibitemOpen
  \bibfield  {author} {\bibinfo {author} {\bibfnamefont {L.}~\bibnamefont
  {Lehner}}\ and\ \bibinfo {author} {\bibfnamefont {F.}~\bibnamefont
  {Pretorius}},\ }\href {\doibase 10.1146/annurev-astro-081913-040031}
  {\bibfield  {journal} {\bibinfo  {journal} {Ann. Rev. Astron. Astrophys.}\
  }\textbf {\bibinfo {volume} {52}},\ \bibinfo {pages} {661} (\bibinfo {year}
  {2014})},\ \Eprint {http://arxiv.org/abs/1405.4840} {arXiv:1405.4840
  [astro-ph.HE]} \BibitemShut {NoStop}%
\bibitem [{\citenamefont {Teukolsky}(1973)}]{Teukolsky_APJ1973}%
  \BibitemOpen
  \bibfield  {author} {\bibinfo {author} {\bibfnamefont {S.~A.}\ \bibnamefont
  {Teukolsky}},\ }\href {\doibase 10.1086/152444} {\bibfield  {journal}
  {\bibinfo  {journal} {Astrophys. J.}\ }\textbf {\bibinfo {volume} {185}},\
  \bibinfo {pages} {635} (\bibinfo {year} {1973})}\BibitemShut {NoStop}%
\bibitem [{\citenamefont {Teukolsky}\ and\ \citenamefont
  {Press}(1974)}]{Teukolsky_APJ1974}%
  \BibitemOpen
  \bibfield  {author} {\bibinfo {author} {\bibfnamefont {S.~A.}\ \bibnamefont
  {Teukolsky}}\ and\ \bibinfo {author} {\bibfnamefont {W.~H.}\ \bibnamefont
  {Press}},\ }\href {\doibase 10.1086/153180} {\bibfield  {journal} {\bibinfo
  {journal} {Astrophys. J.}\ }\textbf {\bibinfo {volume} {193}},\ \bibinfo
  {pages} {443} (\bibinfo {year} {1974})}\BibitemShut {NoStop}%
\bibitem [{\citenamefont {Hannam}\ \emph {et~al.}(2014)\citenamefont {Hannam},
  \citenamefont {Schmidt}, \citenamefont {Boh\'e}, \citenamefont {Haegel},
  \citenamefont {Husa}, \citenamefont {Ohme}, \citenamefont {Pratten},\ and\
  \citenamefont {P\"urrer}}]{Hannam_PRL2015}%
  \BibitemOpen
  \bibfield  {author} {\bibinfo {author} {\bibfnamefont {M.}~\bibnamefont
  {Hannam}}, \bibinfo {author} {\bibfnamefont {P.}~\bibnamefont {Schmidt}},
  \bibinfo {author} {\bibfnamefont {A.}~\bibnamefont {Boh\'e}}, \bibinfo
  {author} {\bibfnamefont {L.}~\bibnamefont {Haegel}}, \bibinfo {author}
  {\bibfnamefont {S.}~\bibnamefont {Husa}}, \bibinfo {author} {\bibfnamefont
  {F.}~\bibnamefont {Ohme}}, \bibinfo {author} {\bibfnamefont {G.}~\bibnamefont
  {Pratten}}, \ and\ \bibinfo {author} {\bibfnamefont {M.}~\bibnamefont
  {P\"urrer}},\ }\href {\doibase 10.1103/PhysRevLett.113.151101} {\bibfield
  {journal} {\bibinfo  {journal} {Phys. Rev. Lett.}\ }\textbf {\bibinfo
  {volume} {113}},\ \bibinfo {pages} {151101} (\bibinfo {year}
  {2014})}\BibitemShut {NoStop}%
\bibitem [{\citenamefont {Arun}(2012)}]{Arun_PRD2012}%
  \BibitemOpen
  \bibfield  {author} {\bibinfo {author} {\bibfnamefont {K.~G.}\ \bibnamefont
  {Arun}},\ }\href {\doibase 10.1088/0264-9381/29/7/075011} {\bibfield
  {journal} {\bibinfo  {journal} {Class. Quant. Grav.}\ }\textbf {\bibinfo
  {volume} {29}},\ \bibinfo {pages} {075011} (\bibinfo {year} {2012})},\
  \Eprint {http://arxiv.org/abs/1202.5911} {arXiv:1202.5911 [gr-qc]}
  \BibitemShut {NoStop}%
\bibitem [{\citenamefont {{LIGO Scientific Collaboration}}(2018)}]{lalsuite}%
  \BibitemOpen
  \bibfield  {author} {\bibinfo {author} {\bibnamefont {{LIGO Scientific
  Collaboration}}},\ }\href {\doibase 10.7935/GT1W-FZ16} {\enquote {\bibinfo
  {title} {{LIGO} {A}lgorithm {L}ibrary - {LALS}uite},}\ }\bibinfo
  {howpublished} {free software (GPL)} (\bibinfo {year} {2018})\BibitemShut
  {NoStop}%
\bibitem [{\citenamefont {Yunes}\ \emph {et~al.}(2016)\citenamefont {Yunes},
  \citenamefont {Yagi},\ and\ \citenamefont {Pretorius}}]{Yunes_PRD2016}%
  \BibitemOpen
  \bibfield  {author} {\bibinfo {author} {\bibfnamefont {N.}~\bibnamefont
  {Yunes}}, \bibinfo {author} {\bibfnamefont {K.}~\bibnamefont {Yagi}}, \ and\
  \bibinfo {author} {\bibfnamefont {F.}~\bibnamefont {Pretorius}},\ }\href
  {\doibase 10.1103/PhysRevD.94.084002} {\bibfield  {journal} {\bibinfo
  {journal} {Phys. Rev. D}\ }\textbf {\bibinfo {volume} {94}},\ \bibinfo
  {pages} {084002} (\bibinfo {year} {2016})}\BibitemShut {NoStop}%
\bibitem [{\citenamefont {Abbott}\ \emph
  {et~al.}(2019{\natexlab{a}})\citenamefont {Abbott}, \citenamefont {Abbott},\
  and\ \citenamefont {Abbott}}]{Test_GR_PRD2019}%
  \BibitemOpen
  \bibfield  {author} {\bibinfo {author} {\bibfnamefont {B.~P.}\ \bibnamefont
  {Abbott}}, \bibinfo {author} {\bibfnamefont {R.}~\bibnamefont {Abbott}}, \
  and\ \bibinfo {author} {\bibfnamefont {T.~D. e.~a.}\ \bibnamefont {Abbott}}
  (\bibinfo {collaboration} {The LIGO Scientific Collaboration and the Virgo
  Collaboration}),\ }\href {\doibase 10.1103/PhysRevD.100.104036} {\bibfield
  {journal} {\bibinfo  {journal} {Phys. Rev. D}\ }\textbf {\bibinfo {volume}
  {100}},\ \bibinfo {pages} {104036} (\bibinfo {year}
  {2019}{\natexlab{a}})}\BibitemShut {NoStop}%
\bibitem [{\citenamefont {Tahura}\ \emph {et~al.}(2019)\citenamefont {Tahura},
  \citenamefont {Yagi},\ and\ \citenamefont {Carson}}]{Tahura_PRD2019}%
  \BibitemOpen
  \bibfield  {author} {\bibinfo {author} {\bibfnamefont {S.}~\bibnamefont
  {Tahura}}, \bibinfo {author} {\bibfnamefont {K.}~\bibnamefont {Yagi}}, \ and\
  \bibinfo {author} {\bibfnamefont {Z.}~\bibnamefont {Carson}},\ }\href
  {\doibase 10.1103/PhysRevD.100.104001} {\bibfield  {journal} {\bibinfo
  {journal} {Phys. Rev.}\ }\textbf {\bibinfo {volume} {D100}},\ \bibinfo
  {pages} {104001} (\bibinfo {year} {2019})},\ \Eprint
  {http://arxiv.org/abs/1907.10059} {arXiv:1907.10059 [gr-qc]} \BibitemShut
  {NoStop}%
\bibitem [{\citenamefont {{Christiansen}}\ \emph {et~al.}(2021)\citenamefont
  {{Christiansen}}, \citenamefont {{Jim{\'e}nez}},\ and\ \citenamefont
  {{Mota}}}]{Christiansen_arxiv2020}%
  \BibitemOpen
  \bibfield  {author} {\bibinfo {author} {\bibfnamefont {{\O}.}~\bibnamefont
  {{Christiansen}}}, \bibinfo {author} {\bibfnamefont {J.~B.}\ \bibnamefont
  {{Jim{\'e}nez}}}, \ and\ \bibinfo {author} {\bibfnamefont {D.~F.}\
  \bibnamefont {{Mota}}},\ }\href {\doibase 10.1088/1361-6382/abdaf5}
  {\bibfield  {journal} {\bibinfo  {journal} {Classical and Quantum Gravity}\
  }\textbf {\bibinfo {volume} {38}},\ \bibinfo {eid} {075017} (\bibinfo {year}
  {2021})},\ \Eprint {http://arxiv.org/abs/2003.11452} {arXiv:2003.11452
  [gr-qc]} \BibitemShut {NoStop}%
\bibitem [{\citenamefont {{Abbott}}\ \emph
  {et~al.}(2021{\natexlab{a}})\citenamefont {{Abbott}}, \citenamefont
  {{Abbott}}, \citenamefont {{Abraham}}, \citenamefont {{Acernese}},
  \citenamefont {{Ackley}}, \citenamefont {{Adams}}, \citenamefont
  {{Adhikari}}, \citenamefont {{Adya}}, \citenamefont {{Affeldt}},
  \citenamefont {{Agathos}},\ and\ \citenamefont
  {et~al.}}]{LIGO_O1O2data_arxiv2019}%
  \BibitemOpen
  \bibfield  {author} {\bibinfo {author} {\bibfnamefont {R.}~\bibnamefont
  {{Abbott}}}, \bibinfo {author} {\bibfnamefont {T.~D.}\ \bibnamefont
  {{Abbott}}}, \bibinfo {author} {\bibfnamefont {S.}~\bibnamefont {{Abraham}}},
  \bibinfo {author} {\bibfnamefont {F.}~\bibnamefont {{Acernese}}}, \bibinfo
  {author} {\bibfnamefont {K.}~\bibnamefont {{Ackley}}}, \bibinfo {author}
  {\bibfnamefont {C.}~\bibnamefont {{Adams}}}, \bibinfo {author} {\bibfnamefont
  {R.~X.}\ \bibnamefont {{Adhikari}}}, \bibinfo {author} {\bibfnamefont
  {V.~B.}\ \bibnamefont {{Adya}}}, \bibinfo {author} {\bibfnamefont
  {C.}~\bibnamefont {{Affeldt}}}, \bibinfo {author} {\bibfnamefont
  {M.}~\bibnamefont {{Agathos}}}, \ and\ \bibinfo {author} {\bibnamefont
  {et~al.}},\ }\href {\doibase 10.1016/j.softx.2021.100658} {\bibfield
  {journal} {\bibinfo  {journal} {SoftwareX}\ }\textbf {\bibinfo {volume}
  {13}},\ \bibinfo {eid} {100658} (\bibinfo {year} {2021}{\natexlab{a}})},\
  \Eprint {http://arxiv.org/abs/1912.11716} {arXiv:1912.11716 [gr-qc]}
  \BibitemShut {NoStop}%
\bibitem [{\citenamefont {{Ashton}}\ \emph {et~al.}(2019)\citenamefont
  {{Ashton}}, \citenamefont {{H{\"u}bner}}, \citenamefont {{Lasky}},
  \citenamefont {{Talbot}}, \citenamefont {{Ackley}}, \citenamefont
  {{Biscoveanu}}, \citenamefont {{Chu}}, \citenamefont {{Divakarla}},
  \citenamefont {{Easter}}, \citenamefont {{Goncharov}}, \citenamefont
  {{Hernandez Vivanco}}, \citenamefont {{Harms}}, \citenamefont {{Lower}},
  \citenamefont {{Meadors}}, \citenamefont {{Melchor}}, \citenamefont
  {{Payne}}, \citenamefont {{Pitkin}}, \citenamefont {{Powell}}, \citenamefont
  {{Sarin}}, \citenamefont {{Smith}},\ and\ \citenamefont
  {{Thrane}}}]{Ashton_APJ2019}%
  \BibitemOpen
  \bibfield  {author} {\bibinfo {author} {\bibfnamefont {G.}~\bibnamefont
  {{Ashton}}}, \bibinfo {author} {\bibfnamefont {M.}~\bibnamefont
  {{H{\"u}bner}}}, \bibinfo {author} {\bibfnamefont {P.~D.}\ \bibnamefont
  {{Lasky}}}, \bibinfo {author} {\bibfnamefont {C.}~\bibnamefont {{Talbot}}},
  \bibinfo {author} {\bibfnamefont {K.}~\bibnamefont {{Ackley}}}, \bibinfo
  {author} {\bibfnamefont {S.}~\bibnamefont {{Biscoveanu}}}, \bibinfo {author}
  {\bibfnamefont {Q.}~\bibnamefont {{Chu}}}, \bibinfo {author} {\bibfnamefont
  {A.}~\bibnamefont {{Divakarla}}}, \bibinfo {author} {\bibfnamefont {P.~J.}\
  \bibnamefont {{Easter}}}, \bibinfo {author} {\bibfnamefont {B.}~\bibnamefont
  {{Goncharov}}}, \bibinfo {author} {\bibfnamefont {F.}~\bibnamefont
  {{Hernandez Vivanco}}}, \bibinfo {author} {\bibfnamefont {J.}~\bibnamefont
  {{Harms}}}, \bibinfo {author} {\bibfnamefont {M.~E.}\ \bibnamefont
  {{Lower}}}, \bibinfo {author} {\bibfnamefont {G.~D.}\ \bibnamefont
  {{Meadors}}}, \bibinfo {author} {\bibfnamefont {D.}~\bibnamefont
  {{Melchor}}}, \bibinfo {author} {\bibfnamefont {E.}~\bibnamefont {{Payne}}},
  \bibinfo {author} {\bibfnamefont {M.~D.}\ \bibnamefont {{Pitkin}}}, \bibinfo
  {author} {\bibfnamefont {J.}~\bibnamefont {{Powell}}}, \bibinfo {author}
  {\bibfnamefont {N.}~\bibnamefont {{Sarin}}}, \bibinfo {author} {\bibfnamefont
  {R.~J.~E.}\ \bibnamefont {{Smith}}}, \ and\ \bibinfo {author} {\bibfnamefont
  {E.}~\bibnamefont {{Thrane}}},\ }\href {\doibase 10.3847/1538-4365/ab06fc}
  {\bibfield  {journal} {\bibinfo  {journal} {\apjs}\ }\textbf {\bibinfo
  {volume} {241}},\ \bibinfo {eid} {27} (\bibinfo {year} {2019})},\ \Eprint
  {http://arxiv.org/abs/1811.02042} {arXiv:1811.02042 [astro-ph.IM]}
  \BibitemShut {NoStop}%
\bibitem [{\citenamefont {{Buchner}}\ \emph {et~al.}(2014)\citenamefont
  {{Buchner}}, \citenamefont {{Georgakakis}}, \citenamefont {{Nandra}},
  \citenamefont {{Hsu}}, \citenamefont {{Rangel}}, \citenamefont {{Brightman}},
  \citenamefont {{Merloni}}, \citenamefont {{Salvato}}, \citenamefont
  {{Donley}},\ and\ \citenamefont {{Kocevski}}}]{Buchner_AAP2014}%
  \BibitemOpen
  \bibfield  {author} {\bibinfo {author} {\bibfnamefont {J.}~\bibnamefont
  {{Buchner}}}, \bibinfo {author} {\bibfnamefont {A.}~\bibnamefont
  {{Georgakakis}}}, \bibinfo {author} {\bibfnamefont {K.}~\bibnamefont
  {{Nandra}}}, \bibinfo {author} {\bibfnamefont {L.}~\bibnamefont {{Hsu}}},
  \bibinfo {author} {\bibfnamefont {C.}~\bibnamefont {{Rangel}}}, \bibinfo
  {author} {\bibfnamefont {M.}~\bibnamefont {{Brightman}}}, \bibinfo {author}
  {\bibfnamefont {A.}~\bibnamefont {{Merloni}}}, \bibinfo {author}
  {\bibfnamefont {M.}~\bibnamefont {{Salvato}}}, \bibinfo {author}
  {\bibfnamefont {J.}~\bibnamefont {{Donley}}}, \ and\ \bibinfo {author}
  {\bibfnamefont {D.}~\bibnamefont {{Kocevski}}},\ }\href {\doibase
  10.1051/0004-6361/201322971} {\bibfield  {journal} {\bibinfo  {journal}
  {\aap}\ }\textbf {\bibinfo {volume} {564}},\ \bibinfo {eid} {A125} (\bibinfo
  {year} {2014})},\ \Eprint {http://arxiv.org/abs/1402.0004} {arXiv:1402.0004
  [astro-ph.HE]} \BibitemShut {NoStop}%
\bibitem [{\citenamefont {Abbott}\ \emph
  {et~al.}(2019{\natexlab{b}})\citenamefont {Abbott}, \citenamefont {Abbott},\
  and\ \citenamefont {Abbott}}]{LIGO_PRX2019}%
  \BibitemOpen
  \bibfield  {author} {\bibinfo {author} {\bibfnamefont {B.~P.}\ \bibnamefont
  {Abbott}}, \bibinfo {author} {\bibfnamefont {R.}~\bibnamefont {Abbott}}, \
  and\ \bibinfo {author} {\bibfnamefont {T.~D. e.~a.}\ \bibnamefont {Abbott}}
  (\bibinfo {collaboration} {LIGO Scientific Collaboration and Virgo
  Collaboration}),\ }\href {\doibase 10.1103/PhysRevX.9.031040} {\bibfield
  {journal} {\bibinfo  {journal} {Phys. Rev. X}\ }\textbf {\bibinfo {volume}
  {9}},\ \bibinfo {pages} {031040} (\bibinfo {year}
  {2019}{\natexlab{b}})}\BibitemShut {NoStop}%
\bibitem [{\citenamefont {{Wang}}\ \emph {et~al.}(2021)\citenamefont {{Wang}},
  \citenamefont {{Tang}}, \citenamefont {{Li}}, \citenamefont {{Han}},\ and\
  \citenamefont {{Fan}}}]{2021arXiv210407590W}%
  \BibitemOpen
  \bibfield  {author} {\bibinfo {author} {\bibfnamefont {H.-T.}\ \bibnamefont
  {{Wang}}}, \bibinfo {author} {\bibfnamefont {S.-P.}\ \bibnamefont {{Tang}}},
  \bibinfo {author} {\bibfnamefont {P.-C.}\ \bibnamefont {{Li}}}, \bibinfo
  {author} {\bibfnamefont {M.-Z.}\ \bibnamefont {{Han}}}, \ and\ \bibinfo
  {author} {\bibfnamefont {Y.-Z.}\ \bibnamefont {{Fan}}},\ }\href@noop {}
  {\bibfield  {journal} {\bibinfo  {journal} {arXiv e-prints}\ ,\ \bibinfo
  {eid} {arXiv:2104.07590}} (\bibinfo {year} {2021})},\ \Eprint
  {http://arxiv.org/abs/2104.07590} {arXiv:2104.07590 [gr-qc]} \BibitemShut
  {NoStop}%
\bibitem [{\citenamefont {{Bozzola}}\ and\ \citenamefont
  {{Paschalidis}}(2021)}]{2021PhRvL.126d1103B}%
  \BibitemOpen
  \bibfield  {author} {\bibinfo {author} {\bibfnamefont {G.}~\bibnamefont
  {{Bozzola}}}\ and\ \bibinfo {author} {\bibfnamefont {V.}~\bibnamefont
  {{Paschalidis}}},\ }\href {\doibase 10.1103/PhysRevLett.126.041103}
  {\bibfield  {journal} {\bibinfo  {journal} {\prl}\ }\textbf {\bibinfo
  {volume} {126}},\ \bibinfo {eid} {041103} (\bibinfo {year} {2021})},\ \Eprint
  {http://arxiv.org/abs/2006.15764} {arXiv:2006.15764 [gr-qc]} \BibitemShut
  {NoStop}%
\bibitem [{\citenamefont {{Amaro-Seoane}}\ \emph {et~al.}(2017)\citenamefont
  {{Amaro-Seoane}}, \citenamefont {{Audley}}, \citenamefont {{Babak}},
  \citenamefont {{Baker}} \emph {et~al.}}]{LISA_arxiv2017}%
  \BibitemOpen
  \bibfield  {author} {\bibinfo {author} {\bibfnamefont {P.}~\bibnamefont
  {{Amaro-Seoane}}}, \bibinfo {author} {\bibfnamefont {H.}~\bibnamefont
  {{Audley}}}, \bibinfo {author} {\bibfnamefont {S.}~\bibnamefont {{Babak}}},
  \bibinfo {author} {\bibfnamefont {J.}~\bibnamefont {{Baker}}},  \emph
  {et~al.},\ }\href@noop {} {\bibfield  {journal} {\bibinfo  {journal} {ArXiv
  e-prints}\ ,\ \bibinfo {eid} {arXiv:1702.00786}} (\bibinfo {year} {2017})},\
  \Eprint {http://arxiv.org/abs/1702.00786} {arXiv:1702.00786 [astro-ph.IM]}
  \BibitemShut {NoStop}%
\bibitem [{\citenamefont {Luo}\ \emph {et~al.}(2016)\citenamefont {Luo} \emph
  {et~al.}}]{TQ_2015}%
  \BibitemOpen
  \bibfield  {author} {\bibinfo {author} {\bibfnamefont {J.}~\bibnamefont
  {Luo}} \emph {et~al.} (\bibinfo {collaboration} {TianQin}),\ }\href {\doibase
  10.1088/0264-9381/33/3/035010} {\bibfield  {journal} {\bibinfo  {journal}
  {Class. Quant. Grav.}\ }\textbf {\bibinfo {volume} {33}},\ \bibinfo {pages}
  {035010} (\bibinfo {year} {2016})},\ \Eprint
  {http://arxiv.org/abs/1512.02076} {arXiv:1512.02076 [astro-ph.IM]}
  \BibitemShut {NoStop}%
\bibitem [{\citenamefont {Liu}\ \emph {et~al.}(2020)\citenamefont {Liu},
  \citenamefont {Hu}, \citenamefont {Zhang},\ and\ \citenamefont
  {Mei}}]{Liu_PRD2020}%
  \BibitemOpen
  \bibfield  {author} {\bibinfo {author} {\bibfnamefont {S.}~\bibnamefont
  {Liu}}, \bibinfo {author} {\bibfnamefont {Y.-M.}\ \bibnamefont {Hu}},
  \bibinfo {author} {\bibfnamefont {J.-d.}\ \bibnamefont {Zhang}}, \ and\
  \bibinfo {author} {\bibfnamefont {J.}~\bibnamefont {Mei}},\ }\href {\doibase
  10.1103/PhysRevD.101.103027} {\bibfield  {journal} {\bibinfo  {journal}
  {Phys. Rev. D}\ }\textbf {\bibinfo {volume} {101}},\ \bibinfo {pages}
  {103027} (\bibinfo {year} {2020})}\BibitemShut {NoStop}%
\bibitem [{\citenamefont {{Wang}}\ \emph {et~al.}(2019)\citenamefont {{Wang}},
  \citenamefont {{Jiang}}, \citenamefont {{Sesana}}, \citenamefont
  {{Barausse}}, \citenamefont {{Huang}}, \citenamefont {{Wang}}, \citenamefont
  {{Feng}}, \citenamefont {{Wang}}, \citenamefont {{Hu}}, \citenamefont
  {{Mei}},\ and\ \citenamefont {{Luo}}}]{WangPRD2019}%
  \BibitemOpen
  \bibfield  {author} {\bibinfo {author} {\bibfnamefont {H.-T.}\ \bibnamefont
  {{Wang}}}, \bibinfo {author} {\bibfnamefont {Z.}~\bibnamefont {{Jiang}}},
  \bibinfo {author} {\bibfnamefont {A.}~\bibnamefont {{Sesana}}}, \bibinfo
  {author} {\bibfnamefont {E.}~\bibnamefont {{Barausse}}}, \bibinfo {author}
  {\bibfnamefont {S.-J.}\ \bibnamefont {{Huang}}}, \bibinfo {author}
  {\bibfnamefont {Y.-F.}\ \bibnamefont {{Wang}}}, \bibinfo {author}
  {\bibfnamefont {W.-F.}\ \bibnamefont {{Feng}}}, \bibinfo {author}
  {\bibfnamefont {Y.}~\bibnamefont {{Wang}}}, \bibinfo {author} {\bibfnamefont
  {Y.-M.}\ \bibnamefont {{Hu}}}, \bibinfo {author} {\bibfnamefont
  {J.}~\bibnamefont {{Mei}}}, \ and\ \bibinfo {author} {\bibfnamefont
  {J.}~\bibnamefont {{Luo}}},\ }\href {\doibase 10.1103/PhysRevD.100.043003}
  {\bibfield  {journal} {\bibinfo  {journal} {\prd}\ }\textbf {\bibinfo
  {volume} {100}},\ \bibinfo {eid} {043003} (\bibinfo {year} {2019})},\ \Eprint
  {http://arxiv.org/abs/1902.04423} {arXiv:1902.04423 [astro-ph.HE]}
  \BibitemShut {NoStop}%
\bibitem [{\citenamefont {Barausse}\ \emph {et~al.}(2016)\citenamefont
  {Barausse}, \citenamefont {Yunes},\ and\ \citenamefont
  {Chamberlain}}]{Barausse_PRL2016}%
  \BibitemOpen
  \bibfield  {author} {\bibinfo {author} {\bibfnamefont {E.}~\bibnamefont
  {Barausse}}, \bibinfo {author} {\bibfnamefont {N.}~\bibnamefont {Yunes}}, \
  and\ \bibinfo {author} {\bibfnamefont {K.}~\bibnamefont {Chamberlain}},\
  }\href {\doibase 10.1103/PhysRevLett.116.241104} {\bibfield  {journal}
  {\bibinfo  {journal} {Phys. Rev. Lett.}\ }\textbf {\bibinfo {volume} {116}},\
  \bibinfo {pages} {241104} (\bibinfo {year} {2016})},\ \Eprint
  {http://arxiv.org/abs/1603.04075} {arXiv:1603.04075 [gr-qc]} \BibitemShut
  {NoStop}%
\bibitem [{\citenamefont {{Kidder}}\ \emph {et~al.}(1993)\citenamefont
  {{Kidder}}, \citenamefont {{Will}},\ and\ \citenamefont
  {{Wiseman}}}]{Kidder:1992fr}%
  \BibitemOpen
  \bibfield  {author} {\bibinfo {author} {\bibfnamefont {L.~E.}\ \bibnamefont
  {{Kidder}}}, \bibinfo {author} {\bibfnamefont {C.~M.}\ \bibnamefont
  {{Will}}}, \ and\ \bibinfo {author} {\bibfnamefont {A.~G.}\ \bibnamefont
  {{Wiseman}}},\ }\href {\doibase 10.1103/PhysRevD.47.R4183} {\bibfield
  {journal} {\bibinfo  {journal} {\prd}\ }\textbf {\bibinfo {volume} {47}},\
  \bibinfo {pages} {R4183} (\bibinfo {year} {1993})},\ \Eprint
  {http://arxiv.org/abs/gr-qc/9211025} {arXiv:gr-qc/9211025 [gr-qc]}
  \BibitemShut {NoStop}%
\bibitem [{\citenamefont {{Cutler}}\ and\ \citenamefont
  {{Flanagan}}(1994)}]{Cutler:1994ys}%
  \BibitemOpen
  \bibfield  {author} {\bibinfo {author} {\bibfnamefont {C.}~\bibnamefont
  {{Cutler}}}\ and\ \bibinfo {author} {\bibfnamefont {{\'E}.~E.}\ \bibnamefont
  {{Flanagan}}},\ }\href {\doibase 10.1103/PhysRevD.49.2658} {\bibfield
  {journal} {\bibinfo  {journal} {\prd}\ }\textbf {\bibinfo {volume} {49}},\
  \bibinfo {pages} {2658} (\bibinfo {year} {1994})},\ \Eprint
  {http://arxiv.org/abs/gr-qc/9402014} {arXiv:gr-qc/9402014 [gr-qc]}
  \BibitemShut {NoStop}%
\bibitem [{\citenamefont {Maggiore}(2007)}]{Maggiore_1900}%
  \BibitemOpen
  \bibfield  {author} {\bibinfo {author} {\bibfnamefont {M.}~\bibnamefont
  {Maggiore}},\ }\href {http://www.oup.com/uk/catalogue/?ci=9780198570745}
  {\emph {\bibinfo {title} {Gravitational Waves. Vol. 1: Theory and
  Experiments}}},\ Oxford Master Series in Physics\ (\bibinfo  {publisher}
  {Oxford University Press},\ \bibinfo {year} {2007})\BibitemShut {NoStop}%
\bibitem [{\citenamefont {Pugliese}\ \emph {et~al.}(2011)\citenamefont
  {Pugliese}, \citenamefont {Quevedo},\ and\ \citenamefont
  {Ruffini}}]{Pugliese_PRD2011}%
  \BibitemOpen
  \bibfield  {author} {\bibinfo {author} {\bibfnamefont {D.}~\bibnamefont
  {Pugliese}}, \bibinfo {author} {\bibfnamefont {H.}~\bibnamefont {Quevedo}}, \
  and\ \bibinfo {author} {\bibfnamefont {R.}~\bibnamefont {Ruffini}},\ }\href
  {\doibase 10.1103/PhysRevD.83.024021} {\bibfield  {journal} {\bibinfo
  {journal} {Phys. Rev. D}\ }\textbf {\bibinfo {volume} {83}},\ \bibinfo
  {pages} {024021} (\bibinfo {year} {2011})}\BibitemShut {NoStop}%
\bibitem [{\citenamefont {{Abbott}}\ \emph
  {et~al.}(2021{\natexlab{b}})\citenamefont {{Abbott}}, \citenamefont
  {{Abbott}}, \citenamefont {{Abraham}}, \citenamefont {{Acernese}},
  \citenamefont {{Ackley}}, \citenamefont {{Adams}}, \citenamefont {{Adams}},
  \citenamefont {{Adhikari}}, \citenamefont {{Adya}}, \citenamefont
  {{Affeldt}},\ and\ \citenamefont {et~al.}}]{LIGO_O3a_arxiv2020}%
  \BibitemOpen
  \bibfield  {author} {\bibinfo {author} {\bibfnamefont {R.}~\bibnamefont
  {{Abbott}}}, \bibinfo {author} {\bibfnamefont {T.~D.}\ \bibnamefont
  {{Abbott}}}, \bibinfo {author} {\bibfnamefont {S.}~\bibnamefont {{Abraham}}},
  \bibinfo {author} {\bibfnamefont {F.}~\bibnamefont {{Acernese}}}, \bibinfo
  {author} {\bibfnamefont {K.}~\bibnamefont {{Ackley}}}, \bibinfo {author}
  {\bibfnamefont {A.}~\bibnamefont {{Adams}}}, \bibinfo {author} {\bibfnamefont
  {C.}~\bibnamefont {{Adams}}}, \bibinfo {author} {\bibfnamefont {R.~X.}\
  \bibnamefont {{Adhikari}}}, \bibinfo {author} {\bibfnamefont {V.~B.}\
  \bibnamefont {{Adya}}}, \bibinfo {author} {\bibfnamefont {C.}~\bibnamefont
  {{Affeldt}}}, \ and\ \bibinfo {author} {\bibnamefont {et~al.}},\ }\href
  {\doibase 10.1103/PhysRevX.11.021053} {\bibfield  {journal} {\bibinfo
  {journal} {Physical Review X}\ }\textbf {\bibinfo {volume} {11}},\ \bibinfo
  {eid} {021053} (\bibinfo {year} {2021}{\natexlab{b}})},\ \Eprint
  {http://arxiv.org/abs/2010.14527} {arXiv:2010.14527 [gr-qc]} \BibitemShut
  {NoStop}%
\end{thebibliography}%
\end{document}